%% file: main.tex
\documentclass[sigconf]{acmart}
\pdfoutput=1 
\AtBeginDocument{%
  \providecommand\BibTeX{{%
    \normalfont B\kern-0.5em{\scshape i\kern-0.25em b}\kern-0.8em\TeX}}}

\usepackage{soul}
\usepackage{graphicx}
\usepackage{booktabs}
\usepackage{float}
\usepackage{overpic}
\usepackage{float}  
\usepackage{subfloat}
\usepackage{stfloats} 
\usepackage[skip=0.5\baselineskip]{caption}
\usepackage{bm}
\usepackage{amsmath}
\usepackage{booktabs}
\usepackage{multirow}
\usepackage{multicol}
\usepackage{enumitem}
\usepackage{subcaption}
\usepackage{threeparttable}
\usepackage{xspace}
\usepackage{color}
\usepackage[normalem]{ulem}
\usepackage{algorithmicx,algorithm}
\usepackage{algpseudocode}
\usepackage{algorithm,algpseudocode}
\usepackage{marvosym}   
\usepackage{colortbl}   
\usepackage{tcolorbox}  
\tcbuselibrary{breakable}   
\usepackage{wasysym}    
\usepackage[figuresright]{rotating} 

\usepackage[edges]{forest}
\usepackage[tikz]{bclogo}
\usepackage[framemethod=tikz]{mdframed}
\tikzset{%
    parent/.style =          {align=center,text width=1.5cm,rounded corners=3pt, line width=0.3mm, fill=gray!10,draw=gray!80},
    child/.style =           {align=center,text width=2cm,rounded corners=3pt, fill=blue!10,draw=blue!80,line width=0.3mm},  
    referenceblock/.style =  {align=center,text width=1.5cm,rounded corners=2pt},
    pretrain/.style =           {align=center,text width=2cm,rounded corners=3pt, fill=blue!10,draw=blue!80,line width=0.3mm},   
    pretrain_work/.style =           {align=center, text width=6cm,rounded corners=3pt, fill=blue!10,draw=blue!0,line width=0.3mm},  
    template/.style =           {align=center,text width=2cm,rounded corners=3pt, fill=red!10,draw=red!80,line width=0.3mm},   
    template_work/.style =           {align=center,text width=6cm,rounded corners=3pt, fill=red!10,draw=red!0,line width=0.3mm},    
    answer/.style =           {align=center,text width=2cm,rounded corners=3pt, fill= cyan!10,draw= cyan!80,line width=0.3mm},   
    answer_work/.style =           {align=center,text width=6cm,rounded corners=3pt, fill= cyan!10,draw= cyan!0,line width=0.3mm}           
}

\makeatletter
\def\@ACM@checkaffil{
    \if@ACM@instpresent\else
    \ClassWarningNoLine{\@classname}{No institution present for an affiliation}%
    \fi
    \if@ACM@citypresent\else
    \ClassWarningNoLine{\@classname}{No city present for an affiliation}%
    \fi
    \if@ACM@countrypresent\else
        \ClassWarningNoLine{\@classname}{No country present for an affiliation}%
    \fi
}
\makeatother

\setcopyright{none}
\renewcommand\footnotetextcopyrightpermission[1]{}
\newcommand{\lqd}[1]{{\color{black} #1}} 

\usepackage{xspace}
\newcommand{\etal}{\emph{et al.}\xspace}
\newcommand{\eg}{\emph{e.g.,}\xspace}
\newcommand{\ie}{\emph{i.e.,}\xspace}

\newcommand{\tabincell}[2]{\begin{tabular}{@{}#1@{}}#2\end{tabular}}

\setcopyright{acmcopyright}
\copyrightyear{2025}
\acmYear{2025}
\acmDOI{XXXXXXX.XXXXXXX}

\acmConference[Conference acronym 'XX]{Make sure to enter the correct
  conference title from your rights confirmation email}{June 03--05,
  2018}{Woodstock, NY}
\acmISBN{978-1-4503-XXXX-X/18/06}

\begin{document}
\begin{sloppypar}   
\title{Large Language Model Enhanced Recommender Systems: \\ A Survey}

\author{Qidong Liu$^{1,2}$, Xiangyu Zhao$^2$ \Letter, Yuhao Wang$^2$, Yejing Wang$^2$, Zijian Zhang$^{3}$, Yuqi Sun$^{1}$ \\
Xiang Li$^{4}$, Maolin Wang$^{2}$, Pengyue Jia$^{2}$, Chong Chen$^{5}$, Wei Huang$^{6}$, Feng Tian$^{1}$ \Letter}
\thanks{\Letter \ \text{Corresponding authors}}
\affiliation{
    \institution{$^1$Xi'an Jiaotong University, 
    $^2$City University of Hong Kong, $^3$Jilin University, \\
    $^4$Nanyang Technological University, 
    $^5$Tsinghua University, $^5$Independent Researcher}
    \country{}
}
\email{{liuqidong, YuqiSun}@stu.xjtu.edu.cn, xianzhao@cityu.edu.hk}
\email{{yejing.wang, yhwang25-c, Morin.wang, jia.pengyue}@my.cityu.edu.hk, zhangzijian@jlu.edu.cn}
\email{xiang002@e.ntu.edu.sg, cstchenc@163.com, hwdzyx@gmail.com, fengtian@mail.xjtu.edu.cn}


\renewcommand{\shortauthors}{Qidong Liu \etal}

\begin{abstract}
  
  Large Language Model (LLM) has transformative potential in various domains, including recommender systems (RS). There have been a handful of research that focuses on empowering the RS by LLM. However, previous efforts mainly focus on LLM as RS, which may face the challenge of intolerant inference costs by LLM. Recently, the integration of LLM into RS, known as \textbf{\underline{LLM}}-\textbf{\underline{E}}nhanced \textbf{\underline{R}}ecommender \textbf{\underline{S}}ystems (LLMERS), has garnered significant interest due to its potential to address latency and memory constraints in real-world applications. This paper presents a comprehensive survey of the latest research efforts aimed at leveraging LLM to enhance RS capabilities. We identify a critical shift in the field with the move towards incorporating LLM into the online system, notably by avoiding their use during inference. Our survey categorizes the existing LLMERS approaches into three primary types based on the component of the RS model being augmented: \textbf{Knowledge Enhancement}, \textbf{Interaction Enhancement}, and \textbf{Model Enhancement}. We provide an in-depth analysis of each category, discussing the methodologies, challenges, and contributions of recent studies. Furthermore, we highlight several promising research directions that could further advance the field of LLMERS.

\end{abstract}



\maketitle

\input{1Introduction}

\input{2Knowledge}

\input{3Interaction}

\input{4Model}

\input{5Applications}
\input{6Future}

\input{7Conclusion}


\bibliographystyle{ACM-Reference-Format}
\bibliography{main}


\end{sloppypar}
\end{document}

%% file: 1Introduction.tex
\section{Introduction}

The large language model has been proven to own unprecedented emergent capabilities in language understanding and reasoning~\cite{zhao2023survey,chang2024survey,wu2024exploring}. In view of the only utilization of collaborative signals by conventional recommender systems~\cite{bobadilla2013recommender,wang2022autofield,wang2024gprec}, supplementing semantic information for RS by the LLM is attractive.
Thus, many works have been proposed to fill the gap between natural language and recommendation, leading to a powerful RS. 

Despite a certain extent of success in adapting LLM to RS, one significant difference between the dialogue system and the recommender system lies in the \textbf{inference latency}. RS often requires low latency for a mass of requests, while the LLM, \eg LLaMA-7B, can only achieve seconds latency for a response. However, many studies focus on using the LLM for recommendation directly at the early stage~\cite{gao2023chat}, which makes it difficult to meet the demands of real-world applications. Recently, more researchers have paid attention to such an issue and commenced diving into the LLM-enhanced RS for practice. Therefore, to boost such a direction, we write this survey to conclude the most up-to-date works. 

In order to clarify the range of this survey, we first give out the definition of LLMERS: \textit{\textbf{The conventional recommender systems are enhanced by LLM via assistance in training or supplementary for data, while no need for LLM inference during the service}}. Though there have been some surveys about LLM for RS, three critical differences exist. 
\textbf{i)} Most current surveys focus on \lqd{how to use the LLM itself as a better RS,} including generative recommendation~\cite{wu2024survey,li2024large,li2024survey} and discriminative recommendation~\cite{lin2023can,zhao2024recommender,chen2024large,huang2024foundation,vats2024exploring,chen2024all}.
By comparison, our survey addresses the LLMERS especially.
\textbf{ii)} LLM for RS is a cutting-edge direction, which develops rapidly. Some surveys~\cite{wu2024survey,lin2023can,zhao2024recommender,chen2024all} do not include the most up-to-date papers. By comparison, more than $60$ works in this survey were published after 2024.
\textbf{iii)} A few surveys have referred to LLM-enhanced RS~\cite{lin2023can,chen2024all}, but they only focus on the enhancement from the aspect of feature engineering. 
In contrast, this survey is the first one to conclude LLMERS from a comprehensive view, including feature and model aspects.

\subsection{Preliminary}

\begin{figure}[t]
\centering
\includegraphics[width=\linewidth]{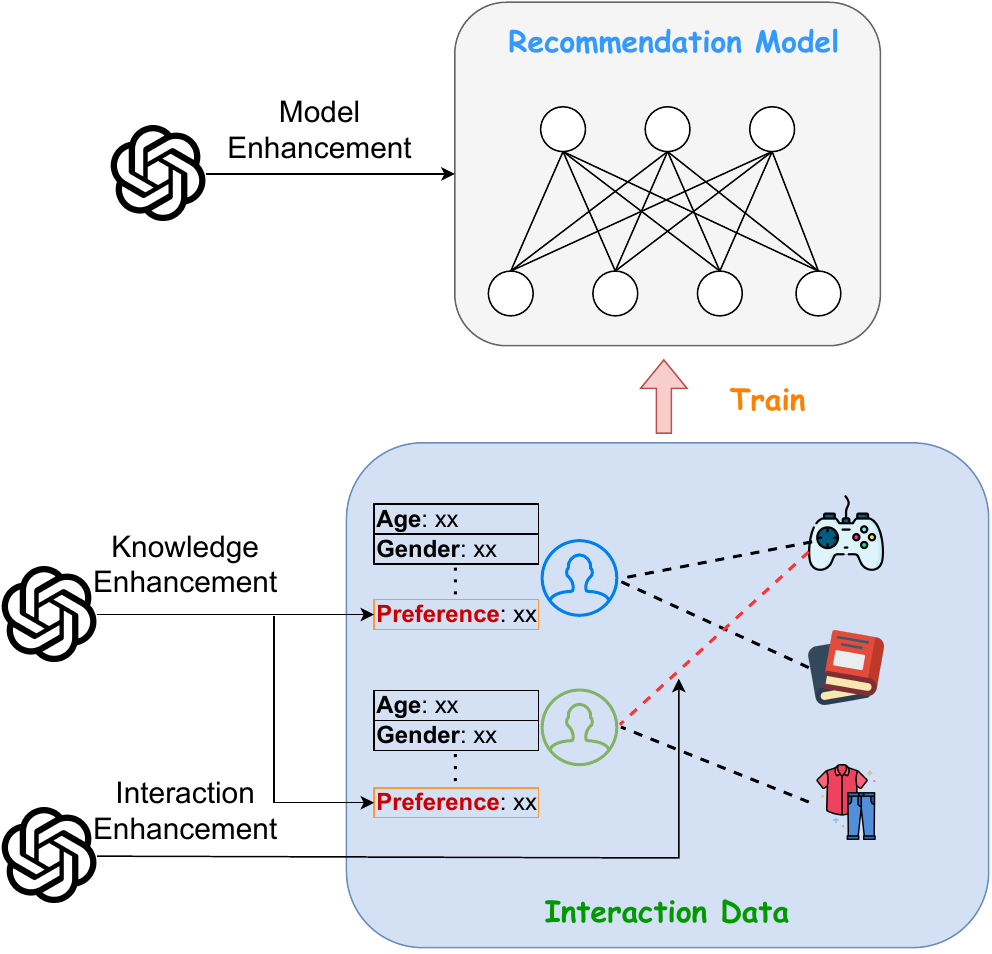}
\caption{The taxonomy of LLM-enhanced RS.}
\label{fig:taxonomy}
\vspace{-5mm}
\end{figure}

Since LLM-enhanced RS targets to enhance RS, introducing RS's components and challenges is a prerequisite to understanding where and why the LLM is needed.
As shown in Figure~\ref{fig:taxonomy}, the conventional RS often consists of the interaction data (feature and interaction) and recommendation model.

\noindent\textbf{Interaction Data}.
The conventional recommender system captures the collaborative signals from user-item records~\cite{koren2021advances}, which depends on the interactions in data for training. Besides, many content-based models~\cite{lops2011content} extract the co-occurrence relationships contained in the features of users and items for recommendation. Therefore, \textbf{Feature} and \textbf{Interaction} are two necessities of data.
However, there exist two challenges of data hindering further advancements for conventional RS. 
\begin{itemize}[leftmargin=*]
    \item \textbf{Challenge 1}: \textit{For features, they are often converted into numerical or categorical values for utilization while lacking reasoning and understanding from the knowledge aspect.}
    \item \textbf{Challenge 2}: \textit{For interactions, data sparsity leads to insufficient training for RS models.}
\end{itemize}

\noindent\textbf{Recommendation Model}.
With the wide application of deep learning techniques, the recommendation model conforms to an ``Embedding-Deep Network'' pattern. The embedding layer targets transforming the raw features into dense representations~\cite{zhao2023embedding}, and the deep network will then capture the user's interests~\cite{zhang2019deep}. However, they both face a unique challenge:
\begin{itemize}[leftmargin=*]
    \item \textbf{Challenge 3}: \textit{For recommendation models, they can only capture the collaborative signals but are unable to leverage semantic information.}
\end{itemize}

\subsection{Taxonomy}
LLMERS augments the conventional RS from its basic components, \ie interaction data and recommendation model, so that only conventional RS models are conducted during serving. 
According to the challenge LLM will take effect, we categorize the LLM-enhanced RS into three lines, which are illustrated in Figure~\ref{fig:taxonomy}.

\noindent\textbf{Knowledge Enhancement}.
This line of work utilizes reasoning abilities and world knowledge of the LLM to derive textual descriptions for users or items. These descriptions will serve as extra features to supplement reasoning and understanding knowledge, and thus they can face the \textbf{Challenge 1}. (Section~\ref{sec:knowledge})

\noindent\textbf{Interaction Enhancement}.
To face the data sparsity issue, \ie \textbf{Challenge 2}, some research studies adopt the LLM to derive new user-item interactions. (Section~\ref{sec:interaction})

\noindent\textbf{Model Enhancement}.
The LLM can analyze the interactions from a semantic view, so some works have tried to make use of the LLM to assist the conventional recommendation models, addressing the \textbf{Challenge 3}. (Section~\ref{sec:model})

\begin{figure*}[!t]
\centering
\begin{forest}
            for tree={
                forked edges,
                grow'=0,
                draw,
                rounded corners,
                node options={align=center,},
                text width=2.7cm,
                s sep=6pt,
                calign=child edge, calign child=(n_children()+1)/2,
            },
            [Prompting Method, fill=gray!45, parent
    [Knowledge \textcolor{blue}{\S\ref{sec:knowledge}}, for tree={pretrain}
        [Summary Text \textcolor{blue}{\S\ref{sec:summary}},  pretrain
            [User Only, pretrain
                [LANE~\cite{zhao2024lane}; LLM-BRec~\cite{jalan2024llmbrec}; RecLM~\cite{jiang2024reclm}
                ,  pretrain_work]
            ]
            [Item Only, pretrain
                [ONCE~\cite{liu2024once}; LAMAR~\cite{luo2024lamar}; 
                GPTAugNews~\cite{yada2024gptaugnews}; SeRALM~\cite{ren2024seralm}; MMRec~\cite{tian2024mmrec}; X-Reflect~\cite{lyu2024xreflect}, pretrain_work]
            ]
            [User \& Item, pretrain
                [KAR~\cite{xi2024kar};  GaCLLM~\cite{du2024gacllm}; SLIM~\cite{wang2024slim}; REKI~\cite{xi2024reki}, pretrain_work]
            ]
        ]
        [Knowledge Graph \textcolor{blue}{\S\ref{sec:kg}}, pretrain
            [Generation, pretrain
                [LLMRG~\cite{wang2024llmrg}; SAGCN~\cite{liu2023sagcn}; LLMHG~\cite{chu2024llmhg};  HRGraph~\cite{wasi2024hrgraph}; LLM-PKG~\cite{wang2024llmpkg}; AutoGraph~\cite{shan2024autograph};
                CIKG~\cite{hu2024cikg}
            , pretrain_work]
            ]
            [Completion \& Fusion, pretrain
                [LLM-KERec~\cite{zhao2024llmkerec}; CSRec~\cite{yang2024csrec}; CoLaKG~\cite{cui2024colakg}, pretrain_work]
            ]
        ]
        [Combination \textcolor{blue}{\S\ref{sec:combination}},  pretrain
            [KELLMRec~\cite{luo2024kellmrec}; SKarRec~\cite{li2024skarrec}, pretrain_work]     
        ]
    ]
    [Interaction \textcolor{blue}{\S\ref{sec:interaction}}, for tree={fill=red!45,template}
        [Text-based \textcolor{blue}{\S\ref{sec:text}},  template
            [ONCE~\cite{liu2024once}; LLMRec~\cite{wei2024llmrec}; LlamaRec~\cite{luo2024llamarec}; ColdAug~\cite{wang2024coldaug}; LLMHD~\cite{song2024llmhd}; LLM4IDRec~\cite{chen2024llm4idrec}; SampleLLM~\cite{gao2025samplellm}
            , template_work]    
        ]
        [Score-based \textcolor{blue}{\S\ref{sec:score}},  template
            [LLM-Ins~\cite{huang2024llmins}; LLM4DSR~\cite{wang2024llm4dsr}; EIMF~\cite{qiao2024eimf}
            , template_work]
        ]
    ]
    [Model \\ \textcolor{blue}{\S\ref{sec:model}}, for tree={fill=blue!45, answer}
        [Model \\ Initialization \textcolor{blue}{\S\ref{sec:init}}, answer
            [Whole, answer
                [CTRL~\cite{li2023ctrl}; FLIP~\cite{wang2024flip}; , answer_work]                            
            ]
            [Embedding, answer
                [LLM2X~\cite{harte2023llm2x}; SAID~\cite{hu2024said}; LEARN~\cite{jia2024learn}; TSLRec~\cite{liu2024tslrec}; LLMEmb~\cite{liu2024llmemb}; LLMInit~\cite{zhang2025llminit}
                , answer_work]             
            ]
        ]                
        [Model \\ Distillation \textcolor{blue}{\S\ref{sec:distill}},  answer
            [Feature, answer
                [LEADER~\cite{liu2024leader}; SLMRec~\cite{xu2024slmrec}, answer_work]  
            ]
            [Response, answer
                [DLLM2Rec~\cite{cui2024dllm2rec}, answer_work]
            ]
        ]
        [Embedding Utilization \textcolor{blue}{\S\ref{sec:utilize}},  answer
            [User Only, answer
                [LLM-CF~\cite{sun2024llmcf}; LLMGPR~\cite{long2024llmgpr}
                , answer_work]    
            ]
            [Item Only, answer
                [TedRec~\cite{xu2024tedrec};  LSVCR~\cite{zheng2024lsvcr}; LRD~\cite{yang2024lrd}; NoteLLM~\cite{zhang2024notellm}; LLM-ESR~\cite{liu2024llmesr}; CELA~\cite{wang2024cela}; NoteLLM-2~\cite{zhang2024notellm2}; \\ AlphaRec~\cite{sheng2024alpharec}; Flare~\cite{hebert2024flare}; HLLM~\cite{chen2024hllm}; QARM~\cite{luo2024qarm}; Molar~\cite{luo2024molar}; PRECISE~\cite{song2024precise}; PAD~\cite{wang2024pad}; RecG~\cite{li2025recg}
                , answer_work]
            ]
            [User \& Item, answer
                [SINGLE~\cite{liu2024single}; 
                LoID~\cite{xu2024loid}; BAHE~\cite{geng2024bahe};  DynLLM~\cite{zhao2024dynllm}; \\ EmbSum~\cite{zhang2024embsum}; Laser~\cite{lin2024laser}; LARR~\cite{wan2024larr}, answer_work]
            ]
        ]
        [Embedding Guidance \textcolor{blue}{\S\ref{sec:guide}},  answer
            [User Only, answer
                [LLM4SBR~\cite{qiao2024llm4sbr}; LLM-ESR~\cite{liu2024llmesr}; \\ LLM4MSR~\cite{wang2024llm4msr}, answer_work]   
            ]
            [User \& Item, answer
                [RLMRec~\cite{ren2024rlmrec}; DaRec~\cite{yang2024darec}; HARec~\cite{ma2024harec}; IRLLRec~\cite{wang2025irllrec}
                , answer_work]
            ]
        ]
    ]    
            ]
        \end{forest}
            \caption{Typology of LLM-Enhanced Recommender Systems.}
\label{fig:typology}
\end{figure*}
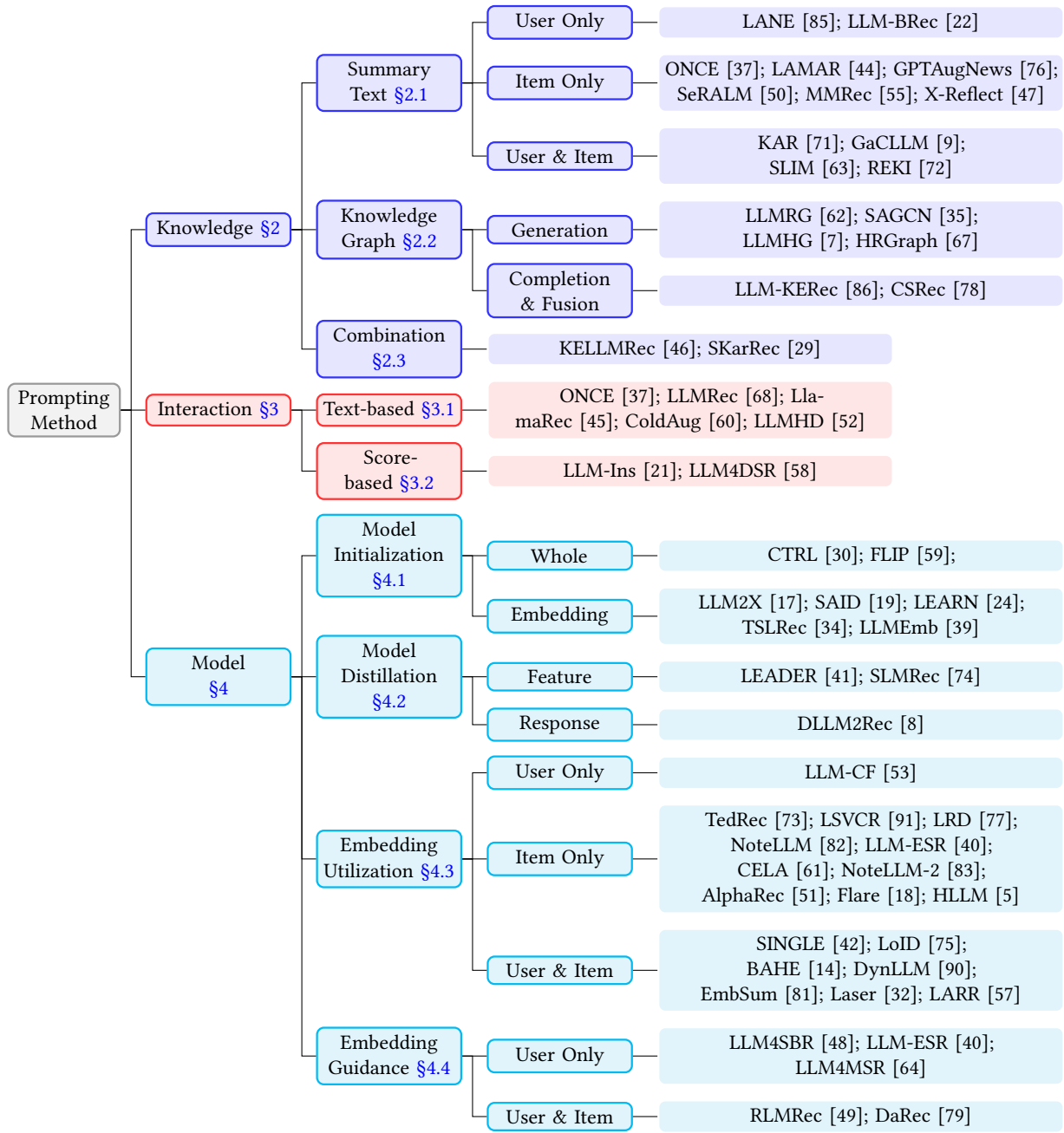

For clarity, we show all the LLMERS papers according to the taxonomy in Figure~\ref{fig:typology}. 



%% file: 2Knowledge.tex
\section{Knowledge Enhancement} \label{sec:knowledge}

\begin{figure}[!t]
\centering
\includegraphics[width=\linewidth]{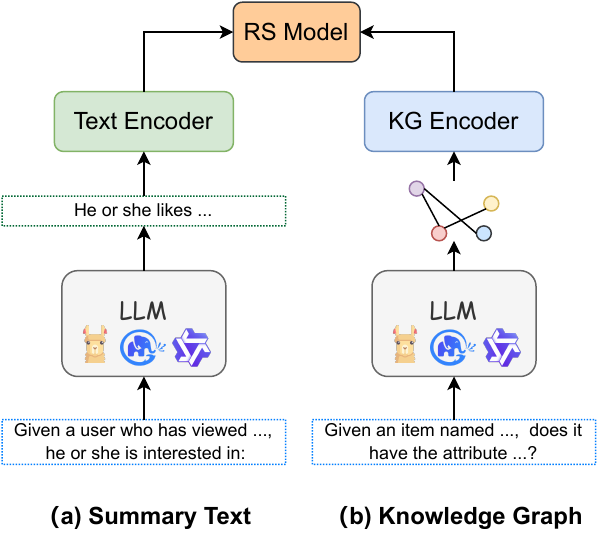}
\caption{The illustration of Knowledge Enhancement.}
\label{fig:knowledge_enhance}
\end{figure}

LLM owns extensive world knowledge and powerful reasoning abilities, which can supplement the RS with external knowledge. In this category, the LLMERS often utilize explicit knowledge to augment the recommender system. For example, as shown in Figure~\ref{fig:taxonomy}, we can prompt the LLM by textual user's interactions to output the ``preference'' of the user as an extra feature. Such a feature, containing the knowledge from the LLM, can enhance the current RS.
\lqd{Since the feature can be derived and saved in advance, the demand for LLM will be avoided during the service, leading to efficient inference.}
According to the type of the feature, we can further partition this line into unstructured \textbf{Summary Text}, structured \textbf{Knowledge graph}, and \textbf{Combination} of these two. The illustrations of these lines are shown in Figure~\ref{fig:knowledge_enhance}.

\subsection{Summary Text} \label{sec:summary}

This subcategory refers to utilizing LLM to summarize the characteristics of items or to reason for the preference of users. For example, the prompt for summarization can be ``Given a user who has viewed $<$\textit{\underline{Browsing History}}$>$, please explain what he or she is interested in: ''. The summary texts are then often encoded by a general language model, such as Bert~\cite{kenton2019bert}.
Some works enhance the RS for either the user or item side, while others address both.

\subsubsection{\textbf{User Only and Item Only}}
For the user-only literature, LANE~\cite{zhao2024lane} and LLM-BRec~\cite{jalan2024llmbrec} fabricate prompts to capture multi-preferences of users. For better utilization, LANE additionally designs an attention mechanism to align the preference representations of RS and LLM. 
On the other hand, as basic units in the RS, more works focus on the item-only aspect. As one of the pioneers in this field, ONCE~\cite{liu2024once} utilizes LLM as a content summarizer for news recommendation. Besides, it proposes to extend the news content by LLM before encoding, while Yada~\etal~\cite{yada2024gptaugnews} uses the LLM to generate category descriptions for augmenting the news content. Similarly, LAMAR~\cite{luo2024lamar} adopts LLM to output factual properties of movies. 
As for SeRALM~\cite{ren2024seralm}, it additionally devises an alignment training pattern for better integration of LLM's knowledge. 
Recently, multimodal RS~\cite{liu2024multimodal} has attracted much attention. MMRec~\cite{tian2024mmrec} and X-Reflect~\cite{lyu2024xreflect} propose to use multimodal LLM to summarize the visual feature into textual description for use.

\subsubsection{\textbf{User \& Item}}
Some studies generate summary texts for both user and item to enhance RS, especially content-based RS~\cite{javed2021review}. As one of the pioneers in this line, KAR~\cite{xi2024kar} gets the reasoning knowledge of users and factual knowledge of items in textual form. 
The knowledge is then encoded by the designed multi-expert text encoders and integrated into the traditional CTR models, such as DeepFM~\cite{guo2017deepfm} and DIN~\cite{zhou2018deep}. 
REKI~\cite{xi2024reki}, a successor of KAR, further refines the knowledge extraction module for various scales of recommendation scenarios. Different from these two works, GaCLLM~\cite{du2024gacllm} enhances the graph-based RS models by the knowledge of users and items, considering the LLM as an aggregator to extract structural information and thereby augment the RS.

\subsection{Knowledge Graph} \label{sec:kg}
Summary text is a type of unstructured knowledge for enhancement, while the structured Knowledge Graph (KG) may drive better integration. Thus, many current works have explored how to apply the LLM to generate a KG (\textbf{Generation}) or augment an existing KG (\textbf{Completion \& Fusion}) for enhancing traditional RS.

\subsubsection{\textbf{Generation}}
In this subcategory, the KG is generated initially and then used as supplementary features for enhancing RS. 
LLMRG~\cite{wang2024llmrg} is the first to explore the KG generation for RS. It fabricates the reasoning prompt to derive possible interaction sequences from LLM and the verification prompt to filter out illogical sequences. 
The generated sequence forms an interaction graph, which is then used to enhance the user representation. 
Similarly, LLMHG~\cite{chu2024llmhg} also prompts LLM to generate a KG based on the user's interaction records, but instead of an interaction graph, it derives user-interest hyper-graphs. 
Besides, SAGCN~\cite{liu2023sagcn} imposes the LLM to distinguish the user-item interactions from various semantic aspects, generating multiple interaction graphs corresponding to these aspects. All of these works tend to generate KG at the interaction level. By comparison, HRGraph~\cite{wasi2024hrgraph} utilizes LLM to conduct job entity extraction from job descriptions and curriculum vitae, producing the specified graph for job recommendation.

\subsubsection{\textbf{Completion \& Fusion}}
Different from generation, this line of work often imposes the LLM on existing KG. For example, LLM-KERec~\cite{zhao2024llmkerec} adopts the LLM to identify complementary relationships within an item KG. 
Then, the KG-based RS can derive better recommendations by enriched entity relationships.
Besides, Yang~\etal~\cite{yang2024csrec} find that the direct fusion of existing KG and LLM-generated entity relations is sub-optimal. To address such an issue, they propose an improved fusion method to better integrate the two sources.

\subsection{Combination} \label{sec:combination}
In consideration of the effectiveness of these two types, some works also resort to combining both of them. KELLMRec~\cite{luo2024kellmrec} is one representative, which integrates the KG for constructing prompts to avoid the hallucination problem when getting the summary texts from LLM. 
Besides, SKarRec~\cite{li2024skarrec} also links the entity in KG to the textual prompts, allowing the LLM to generate KG-aware knowledge that augments the concept recommendations.

%% file: 3Interaction.tex
\section{Interaction Enhancement} \label{sec:interaction}

\begin{figure}[!t]
\centering
\includegraphics[width=\linewidth]{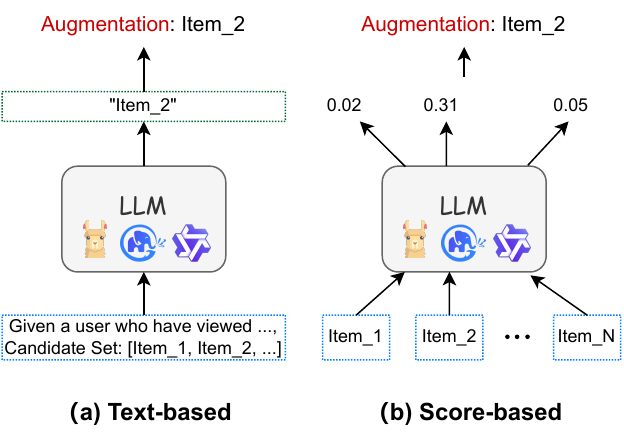}
\caption{The illustration of Interaction Enhancement.}
\label{fig:interaction_enhance}
\end{figure}

To address the sparsity problem in traditional recommender systems, data augmentation is a straightforward way~\cite{lai2024survey}, which targets generating new interactions.
\lqd{The augmented interactions are only supplemented for training traditional RS models, so no extra inference burden will be brought.}
Inspired by the semantic view of LLM, some research studies have explored the interaction generation by LLM. 
We categorize these approaches into two types based on the output of the LLM, \ie \textbf{Text-based} and \textbf{Scored-based}, shown in Figure~\ref{fig:interaction_enhance}. The former gives out the names of pseudo-interacted items as the augmentation. By comparison, scored-based methods derive the logits of the possible interactions and then generate the augmented items by ranking.

\subsection{Text-based} \label{sec:text}

ONCE~\cite{liu2024once} is the first to investigate the data augmentation for RS using LLM. It prompts the LLM to recommend news based on a user's historical records. 
The generated news is added to the dataset as an augmented interaction for this user. 
However, the direct choice from a large item set is difficult for LLM. Thus, later works propose narrowing the selection to a smaller set of items for higher augmentation accuracy. For example, LLMRec~\cite{wei2024llmrec} first adopts well-trained traditional RS to filter out a list of items and then prompt LLM to choose from them. Similarly, LlamaRec~\cite{luo2024llamarec} imposes the LLM to generate interactions from a random-sampled item set. Furthermore, Wang~\etal~\cite{wang2024coldaug} proposes to utilize the LLM to judge which item a user prefers, generating pair-wise interactions for cold-start RS. 
In addition to generating new data, distinguishing noisy data is also a vital topic in data augmentation. 
LLMHD~\cite{song2024llmhd} addresses this issue by applying the LLM to distinguish hard or noisy samples in sequential recommendation.

\subsection{Score-based} \label{sec:score}

As mentioned in the last subsection, it is difficult for LLM to select from amounts of items. Besides, the textual output often faces the out-of-corpus problem~\cite{bao2023bi}, \ie the generated item name cannot correspond to anyone in the item set. To avoid these two problems, some works explore the score-based method. One typical work of this line is LLM-InS~\cite{huang2024llmins}. It firstly imposes the LLM to derive the semantic embeddings of users and items, and then generate augmentations by calculating the similarity between them. By comparison, LLMDSR~\cite{wang2024llm4dsr} considers the probability of word tokens as the confidence in deciding a noisy item and generates the new item by the similarity of embeddings.

%% file: 4Model.tex
\section{Model Enhancement} \label{sec:model}

\begin{figure}[!t]
\centering
\includegraphics[width=\linewidth]{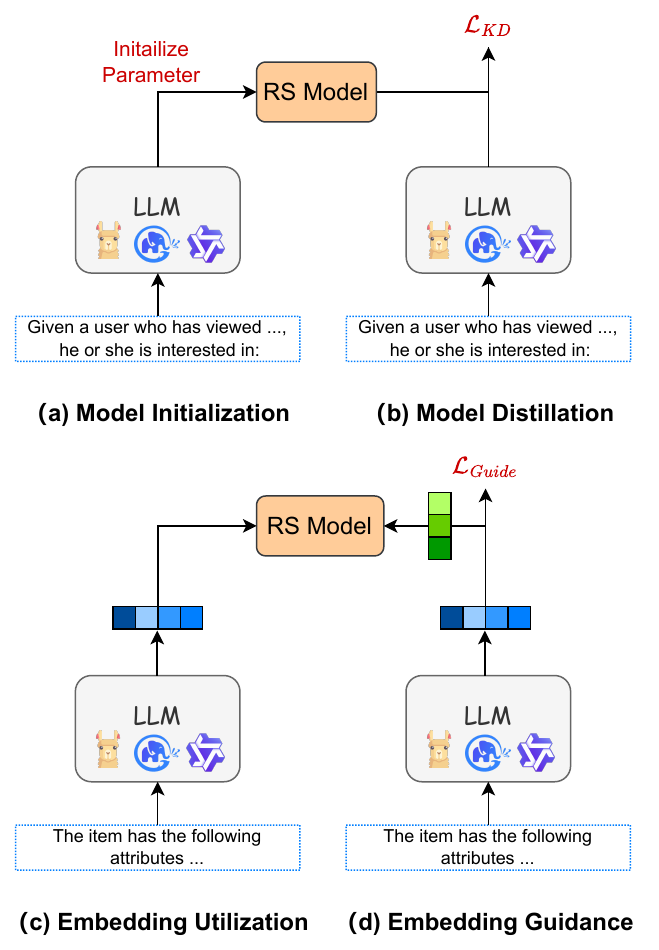}
\caption{The illustration of Model Enhancement.}
\label{fig:model_enhance}
\end{figure}

Beyond data-level enhancement, the powerful abilities and semantics of LLM can also be injected into traditional recommendation models directly. 
RS models are generally composed of two main parts, \ie embedding layer and hidden layers. The former often captures relationships between items, while the latter extracts more complex user preferences. According to the various parts, we categorize this line into \textbf{Model Initialization}, \textbf{Model Distillation}, \textbf{Embedding Utilization} and \textbf{Embedding Guidance}, as shown in Figure~\ref{fig:model_enhance}. 
The first two categories involve applying the LLM to the entire RS model, while embedding utilization and guidance only impose LLM on the embedding layer. 
\lqd{It is worth noting that Model Initialization, Model Distillation, and Embedding Guidance only need the LLM to assist in training RS models, so LLM usage is eliminated during serving. Besides, Embedding Utilization can cache the LLM embeddings before inference. Thus, they all can meet the latency requirements of real-time service.}

\subsection{Model Initialization} \label{sec:init}

Pretrain, as one of the initialization techniques, has shown potential in the recommendation field~\cite{liu2023pre}. 
On the one hand, pretrained weights can help model convergence. On the other hand, the information learned from the pretrain stage will preserve and affect the downstream recommendation tasks. Thus, recent works have explored saving the semantics of LLM to the weights of RS models before training. 
Some approaches apply pretrained weights to the whole RS model, denoted as \textbf{Whole}, while the others are for the embedding layer, referred to as \textbf{Embedding}.

\subsubsection{\textbf{Whole}}
CTRL~\cite{li2023ctrl} and FLIP~\cite{wang2024flip} are the representatives in this line. 
CTRL first considers the semantic representation extracted from LLM and collaborative representation derived from the traditional RS model as two distinct modalities~\cite{fang2022multi,fang2023annotations,fang2024fewer}. 
It then aligns such two ``modalities'' by contrastive tasks. The parameters updated during the alignment process serve as the initialization of the whole traditional RS model. 
Building upon CTRL, FLIP~\cite{wang2024flip} fabricates fine-grained masked tabular and language modeling alignment tasks for better model initialization.

\subsubsection{\textbf{Embedding}}
Embedding plays a key role in the traditional RS models~\cite{zhao2023embedding}, preliminarily capturing the relationships between items and features. The LLMERS in this sub-category all aim to extract semantic relationships between items by the derived embeddings from LLM (\ie LLM embeddings). 
Harte~\etal~\cite{harte2023llm2x} initiates this line, who input the item title to a large text embedding model. Then, the derived LLM embeddings are used to initialize the embedding layer of a sequential recommendation model. Later, LEARN~\cite{jia2024learn} proposes to adopt the hidden states of a frozen open-sourced LLM as the item embedding. 
However, the general-purpose LLM without fine-tuning may encounter incompatible challenges for the recommendation task. To address this, SAID~\cite{hu2024said} and TSLRec~\cite{liu2024tslrec} specialize the item attribute word generation and information reconstruction tasks, respectively, to adapt the LLM to recommendation better. 
More recently, LLMEmb~\cite{liu2024llmemb} integrates an attribute-level contrastive task to further extract fine-grained semantic relationships between items. 

\subsection{Model Distillation} \label{sec:distill}

Knowledge distillation (KD) is one of the compression techniques~\cite{gou2021knowledge}, which is also promising to bring the powerful abilities of LLM to small RS models. However, the gap between discriminative RS models and generative LLM may hinder the success of KD. 
Thus, existing studies design \textbf{Feature-based} and \textbf{Response-based} distillation ways to address such an issue.

\subsubsection{\textbf{Feature-based}}
Feature-based distillation transfers knowledge directly between the hidden states of models, regardless of the output type. LEADER~\cite{liu2024leader} is the pioneer of this line, which distills the RS model using the hidden state from the final layer of the LLM with a trainable adapter. Notably, LEADER also fine-tunes the LLM to transform the LLM into a discriminative model. In contrast, SLMRec~\cite{xu2024slmrec} maintains the basic structure of LLM, \ie stack of several transformer layers, but reduces the number of layers to create a smaller-scale RS model. It then imposes knowledge distillation to several transformer layers between LLM and the RS model.

\subsubsection{\textbf{Response-based}}
Following the advancement in KD for recommendation, recent works attempt to use LLM to generate a ranking list from a semantic view, followed by the ranking distillation loss~\cite{tang2018ranking} for the RS model. 
Specifically, DLLM2Rec~\cite{cui2024dllm2rec} proposes to fine-tune the LLM to adapt the recommendation task and then distill the small RS model by the ranking list derived from the LLM.

\subsection{Embedding Utilization}  \label{sec:utilize}
In general, the LLM generates natural language to complete various tasks. However, the textual outputs are often difficult and inefficient to be integrated into RS models directly. 
To avoid the deficiency and better utilize the semantics of LLM, many recent works utilize the embeddings derived from LLM, \eg the hidden states of the last transformer layer of the LLM, to enhance traditional RS as a semantic supplementary. From the aspect that LLM embeddings address in the traditional RS, this line of works can be categorized into \textbf{User Only}, \textbf{Item Only} and \textbf{User \& Item}. 

\subsubsection{\textbf{User Only}}
In this cluster, the LLM is used to generate the representation of the user's preference directly. 
The LLM embeddings then will be fed into the traditional RS model for augmentation. A notable example in this line is LLM-CF~\cite{sun2024llmcf}. It first designs a data mixture method to fine-tune the LLM for recommendation capability. Next, the well-trained LLM can generate Chain-of-Thought (CoT) reasoning to enhance user preferences.

\subsubsection{\textbf{Item Only}}
In some types of recommender systems, \eg sequential recommendation, the item embeddings play a fundamental role. Thus, most existing embedding utilization focuses on the item-only enhancement. Many early studies in this sub-category directly adopt the frozen LLM without fine-tuning. 
For example, TedRec~\cite{xu2024tedrec} applies LLaMA to encode the item texts and then proposes a frequency-based fusion to combine the derived semantic embeddings and ID embeddings. 
LRD~\cite{yang2024lrd} adopts LLM embeddings akin to TedRec but introduces a novel self-supervised relation discovery task to optimize the utilization. However, these works may lose the informative semantics of LLM embeddings during training. To address such issue, LLM-ESR~\cite{liu2024llmesr} and AlphaRec~\cite{sheng2024alpharec} both devise a new pattern of frozen LLM embeddings with a trainable adapter to maintain the original semantic information. 

Despite their successful utilization, the general-purpose LLM is often not well suited for the recommendation tasks. 
Thus, several works propose to fine-tune the LLM before utilization.
LEARN~\cite{jia2024learn} proposes to align the user preferences derived from LLM with those of RS models, which refines the item LLM embeddings. NoteLLM~\cite{zhang2024notellm} and NoteLLM-2~\cite{zhang2024notellm2} are both devised for item-item recommendation, where the LLM derives the item embeddings for retrieval. NoteLLM only targets combining the textual semantics and collaborative signals, while NoteLLM-2 gives out a novel path to fine-tune the LLM for understanding the multimodal features of items. Recently, HLLM~\cite{chen2024hllm} proposes a novel hierarchal framework for enhancement. The first level involves an item LLM, which encodes the textual features of the item. 
The item LLM embeddings will then be fed into a user LLM for final recommendation. 

\subsubsection{\textbf{User \& Item}}
In this line, the current papers can be clustered into two categories. One is to utilize the user and item LLM embeddings for recommendation by embedding similarity function directly. For example, LoID~\cite{xu2024loid} first aligns the LLM embeddings with collaborative ID information, then concatenates the user and item representations derived from the LLM for final recommendation. 
However, the user LLM embeddings are often sub-optimal because of over-length textual prompts of the user's interaction histories. To face this challenge, BAHE~\cite{geng2024bahe} proposes to partition an LLM into lower layers and higher layers. It uses the lower layers of the LLM to encode the item attributes to get the comic item representation. Then, the corresponding comic vectors of the user's interacted items are fed into the higher layer. By comparison, EmbSum~\cite{zhang2024embsum} partitions the user's interaction records into several sessions and then uses the LLM to encode them separately. Except for the lengthy prompt issue, the general user LLM embedding also faces a lack of dynamics. To enhance dynamic graph recommendation, DynLLM~\cite{zhao2024dynllm} generates multi-faceted profiles for users. 

The other line of work supplements the user and item LLM embeddings as extra features to augment the content-based RS models. For instance, Laser~\cite{lin2024laser} utilizes the derived user and item LLM embeddings with an adapter for feature enhancement. LARR~\cite{wan2024larr} adopts the LLM embeddings as extra features but further highlights the importance of fine-tuning the LLM specialized for RS.

\subsection{Embedding Guidance} \label{sec:guide}
Different from the direct embedding utilization, this line of papers only uses the LLM embeddings as the guidance for training or parameter synthesis. According to which aspect the LLM embeddings aim to enhance, we further categorize the current works into \textbf{User Only} and \textbf{User \& Item}.

Due to the lengthy prompt of user interactions, user LLM embeddings are often noisy. Thus, considering the LLM embeddings as guidance is a feasible way for \textbf{user-only} enhancement.
LLM4SBR~\cite{qiao2024llm4sbr} enhances the short-term and long-term preferences for session-based recommendation models by identifying the representative items via LLM embeddings. By comparison, LLM-ESR~\cite{liu2024llmesr} retrieves similar users by LLM embeddings to enhance the long-tail users during the training. These two works both aim to search for relevant users or items from a semantic view and enhance the training process. Differently, LLM4MSR~\cite{wang2024llm4msr} generates the user-specified parameters for multi-scenario recommendation models by LLM embeddings accompanied by meta networks.
As for \textbf{User \& Item} line, the research studies guide both the user and item representation learning. RLMRec~\cite{ren2024rlmrec} kicks off this line, which adds an extra loss to align the collaborative user and item embeddings with corresponding LLM embeddings during training. 
Based on RLMRec's insights, DaRec~\cite{yang2024darec} proposes to disentangle the shared and specific parts for both LLM and collaborative embeddings. Only imposing the alignment on shared parts will lead to better performance.

%% file: 5Applications.tex
\section{Trend} \label{sec:appendix_trend}

\begin{figure*}[t]
\centering
\includegraphics[width=\linewidth]{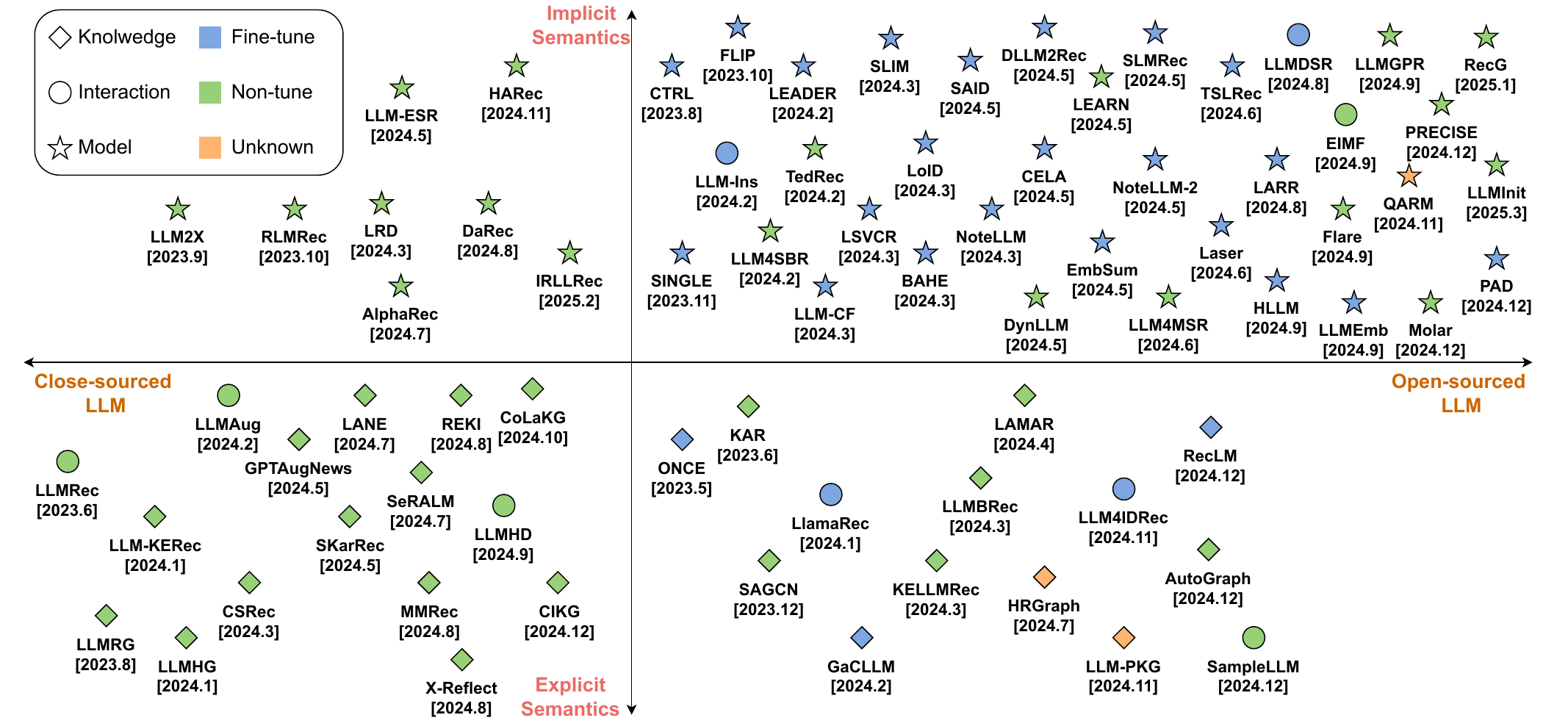}
\caption{The trend of LLM-enhanced recommender systems. The horizontal and vertical axes represent open-sourced LLM or close-sourced LLM and explicit or implicit semantics, respectively. The dots in different colors and shapes are the category of the works and whether the LLM is fine-tuned. In each quadrant, the works are ordered by time from left to right.}
\label{fig:quadrant}
\vspace{-2mm}
\end{figure*}

\begin{table}[!t]
\centering
\caption{The type of LLM used in LLMERS.}
\label{tab:llm_type}
\resizebox{0.5\textwidth}{!}{
\begin{tabular}{cccc}
\toprule
\textbf{Type} & \textbf{Series} & \textbf{Finetune} & \textbf{Related Works} \\ \midrule
\multirow{14}{*}{Open} 
& \multirow{2}{*}{LLaMA} & Y & \tabincell{c}{\cite{liu2024once},\cite{luo2024llamarec}, \cite{huang2024llmins},\cite{wang2024llm4dsr} \\ \cite{liu2024leader},\cite{sun2024llmcf}, \cite{wang2024slim},\cite{zhang2024notellm}\\ \cite{hu2024said},\cite{cui2024dllm2rec}, \cite{xu2024slmrec},\cite{liu2024tslrec} \\ \cite{liu2024llmemb}, \cite{zhang2024notellm2}, \cite{chen2024hllm} }\\
&   & N    &  \tabincell{c}{\cite{luo2024lamar}, \cite{jalan2024llmbrec},\cite{ren2024seralm},\cite{xu2024tedrec} }  \\ 
\cmidrule{2-4} 
& \multirow{2}{*}{ChatGLM}  & Y &  \tabincell{c}{\cite{li2023ctrl},\cite{liu2024single},\cite{zheng2024lsvcr} }             \\
& & N   & \tabincell{c}{\cite{xi2024kar}, \cite{du2024gacllm}, \cite{wang2024llm4msr} }  \\ 
\cmidrule{2-4} 
& \multirow{2}{*}{Qwen}  & Y & \tabincell{c}{\cite{geng2024bahe}, \cite{zhang2024notellm2} }    \\
                &   & N & \tabincell{c}{\cite{qiao2024llm4sbr},\cite{zhao2024dynllm} }        \\ 
\cmidrule{2-4} 
& \multirow{2}{*}{Vicuna} & Y & \cite{wang2024cela}   \\
                &  & N &  \cite{liu2023sagcn}    \\ 
\cmidrule{2-4} 
& \multirow{2}{*}{Mistral}  & Y & \cite{wang2024cela}  \\
                & & N & \cite{zhang2024embsum}    \\ 
\cmidrule{2-4} 
& \multirow{2}{*}{Baichuan} & Y & \tabincell{c}{\cite{wan2024larr}, \cite{chen2024hllm} } \\
                & & N & \tabincell{c}{\cite{jia2024learn} }   \\ 
\cmidrule{2-4} 
& \multirow{2}{*}{InternLM} & Y &  -  \\
                &  & N &  \cite{luo2024kellmrec}, \cite{lyu2024xreflect}             \\ 
\midrule
\multirow{4}{*}[-10mm]{Close} 
& GPT-4 & - &  
    \tabincell{c}{\cite{wang2024llmrg}, \cite{chu2024llmhg}, \cite{yada2024gptaugnews} }  \\ 
\cmidrule{2-4} 
& GPT-3.5 & - &   
    \tabincell{c}{\cite{liu2024once}, \cite{wang2024llmrg}, \cite{chu2024llmhg}, \cite{zhao2024llmkerec} \\ \cite{yang2024csrec}, \cite{li2024skarrec}, \cite{zhao2024lane}, \cite{tian2024mmrec} \\ \cite{xi2024reki}  , \cite{wei2024llmrec}, \cite{song2024llmhd}, \cite{wei2024llmrec} \\ \cite{ren2024rlmrec}, \cite{yang2024darec} }
\\ \cmidrule{2-4} 
& text-embedding-ada & - & \tabincell{c}{\cite{harte2023llm2x},\cite{yang2024lrd}, \cite{liu2024llmesr},\cite{sheng2024alpharec} \\ \cite{wei2024llmrec}, \cite{ren2024rlmrec}, \cite{yang2024darec}  }       \\ \cmidrule{2-4} 
& PaLM  & - &
    \tabincell{c}{\cite{wang2024coldaug} } \\ 
\bottomrule
\end{tabular}
}
\vspace{-2mm}
\end{table}

To investigate the trend of the LLMERS, we visualize the current works based on the type of LLM and semantics in Figure~\ref{fig:quadrant}. As mentioned before, the key to LLM enhancement lies in the semantics. At the early stage, many works~\cite{xi2024kar,zhao2024llmkerec} prompt the LLM to derive the natural language, which contains knowledge and semantics for augmentation. Since they are readable and understandable, we denote them as explicit semantics. By comparison, the cluster of implicit semantics means that the mediation from LLM for enhancement is non-verbal, \eg the hidden states of LLM. Though humans may favor explicit knowledge due to its explainability, we find that recent works prefer the implicit one. The reason lies in the better performance of implicit knowledge while explicit one suffers from information loss during the extra encoding process.

As for the type of LLM, the figure reveals that more LLMERS works resort to the open-sourced LLM. One reason is that it can save money spent on calling the API. More importantly, the open-sourced LLM can be fine-tuned to suit the recommendation task~\cite{liu2024llmemb}. Furthermore, observing the quadrants I and IV, the fine-tuned LLM has become prevalent recently, which indicates the significance of aligning the LLM with recommendation tasks. In Table~\ref{tab:llm_type}, we also conclude the specific type of LLM used in LLMERS.

Overall, we can sum up the trends of current LLMERS as follows:
\begin{itemize}[leftmargin=*]
    \item The current LLMERS studies move from explicit semantics to \textbf{implicit semantics}.
    \item For a better adaptation to the recommendation task, \textbf{fine-tuned open-sourced LLM} become more prevalent.
    \item \textbf{Model enhancement} has gained more attention recently, because it can combine implicit semantics and fine-tuned LLM.
\end{itemize}

\section{Applications, Efficiency, and Resources} \label{sec:appendix_application}

\vspace{1mm}
\textbf{\textit{Applications}}.
Traditional recommender systems have been widely adopted in various applications, such as news recommendations. Though LLMERS is also based on traditional RS, it often requires abundant side information, especially informative texts, to make full use of the reasoning and understanding abilities of LLM. Thus, current LLMERS works often focus on applications that have one of two following characteristics: (i) it has a number of normal \textbf{features} that LLM can understand; (ii) it has a volume of \textbf{texts} that LLM can summarize. The E-commerce application belongs to the former one, where users have many profile features and commodities have attributes~\cite{wang2024llmrg,liu2024llmesr}. Since the LLM owns open-world knowledge, these features can be understood by it and derive semantics for enhancement. By comparison, the news recommendation is the typical application that utilizes the LLM to summarize the abundant news texts~\cite{liu2024once}. To show it more clearly, we list the typical applications that LLMERS has targeted in Table~\ref{tab:dataset}.

\vspace{1mm}
\noindent\textbf{\textit{Efficiency}}.
Compared with LLM as RS, the most important distinction and merit of LLMERS lie in high efficiency. Thus, we discuss how each category of LLMERS is efficient in the paragraph.
i) \textbf{Knowledge Enhancement} minimizes runtime LLM dependency by pre-generating static intermediaries (text summaries or knowledge graphs), but introduces persistent encoding overhead. Summary-based methods append cached summaries as supplementary features via text encoders~\cite{li2023ctrl}, while KG follows dual paths: feature-based KG integrates with content-based systems through embeddings~\cite{huang2024llmins}, and interaction-based KG enhance collaborative filtering via graph processing~\cite{wang2024llmrg}. 
Hybrid methods combine KG-refined summaries with text encoding, trading compounded costs for semantic enrichment.
ii) \textbf{Interaction Enhancement} exclusively amplifies training data through LLM-synthesized user interactions~\cite{yang2024csrec}, preserving inference efficiency and achieving universal scalability across recommendation architectures.
iii) \textbf{Model Enhancement} employs three strategies with varying architectural dependencies:
Initialization of components (content-based models~\cite{li2023ctrl} or embedding layers~\cite{hu2024said}) are enhanced by LLM-generated knowledge, and only small RS models are used for service.
Distillation prioritizes traditional RS models~\cite{cui2024dllm2rec}, eliminating LLM while inference.
Embedding Fusion decouples LLMs from inference by caching embeddings as semantic features (content-based RS~\cite{liu2024single}) or cross-modal alignment anchors (collaborative filtering~\cite{jia2024learn}), incurring only storage and view-encoding costs.

\vspace{1mm}
\noindent\textbf{\textit{Resources}}.
Datasets and benchmarks are often the most vital resources for one research field. For the dataset, many have included either informative features or abundant texts for LLM usage. We list the available public datasets with their access links according to the applications in Table~\ref{tab:dataset} to facilitate further research. As for the benchmark, unfortunately, there has been no one since it is an emerging direction. Nevertheless, many research studies have open-sourced their code, which pushes the advancements of this field. We conclude all the available codes in our survey repository~\footnote{https://github.com/liuqidong07/Awesome-LLM-Enhanced-Recommender-Systems}. 

%% file: 6Future.tex
\section{Future Directions}

LLMERS is a cutting-edge direction, where many problems have not been addressed. Therefore, we give out several directions to inspire later research.

\begin{itemize}[leftmargin=*]
    \item \textbf{Exploration for More Recommendation Tasks}. Existing LLMERS works have focused on revolutionizing several basic recommendation tasks, such as collaborative filtering~\cite{sun2024llmcf} and sequential recommendation~\cite{liu2024llmesr}. They have proven the effectiveness of bringing the semantics of LLM to traditional RS. Thus, it is also promising to replicate their success in some other recommendation tasks.

    \item \textbf{Multimodal RS}. The multimodal RS~\cite{liu2024multimodal} has become prevalent recently due to the emergence of more multimedia services. However, the existing multimodal RS often faces the challenge of extraction and fusion of features in different modalities. 
    The adoption of multimodal LLM~\cite{yin2024survey} may be a feasible way to augment existing multimodal RS and even eliminate the extraction and fusion procedures.

    \item \textbf{User-side Enhancement}. There have been many efforts of LLMERS imposed on the item side, but few on the user side.
    The reason lies in that the user's historical interactions are often organized into texts orderly. However, such a prompt often faces the overlength problem~\cite{geng2024bahe}. 
    Besides, the textual interactions are often difficult to understand for LLM. This problem thus hinders the user-side enhancement severely and needs to be addressed.

    \item \textbf{Explainability}. Explainability is an important aspect of constructing a trustworthy RS. The traditional RS is often short in this aspect since they only adopt meaningless identities. The LLM enhances the RS by understanding users and items from a semantic view, which is promising to derive an explanation. 

    \item \textbf{Benchmark}. Since LLMERS is a newborn direction, there is no benchmark in this field. Therefore, developing a comprehensive and usable benchmark is an urgent need, because it can facilitate fresh researchers to contribute to this field and accelerate the advancement of this direction.
    
\end{itemize}

\begin{table}[!t]
\centering
\caption{The applications and corresponding datasets of LLMERS.}
\resizebox{0.5\textwidth}{!}{
\begin{tabular}{cccc}
\toprule
\textbf{Characteristic} & \textbf{Applications} & \textbf{Dataset} & \textbf{Link} \\ \midrule
\multirow{3}{*}[-20mm]{Features} 
& \multirow{3}{*}{E-commerce} & Amazon & https://bit.ly/4ibCNY1   
      \\ \cmidrule{3-4}
&                             & Alibaba & https://bit.ly/4itdGjq
      \\ \cmidrule{3-4}
&                             & Online Retail & -
      \\ \cmidrule{2-4} 
& \multirow{2}{*}{Movie}  & MovieLens & https://bit.ly/3B6pV4J
      \\ \cmidrule{3-4}
&        & Netflix & https://bitly.is/3ZrfWyT     
      \\ \cmidrule{2-4} 
& \multirow{3}{*}{POI}  & Yelp & https://bit.ly/4gfqvfd 
      \\ \cmidrule{3-4}
&        & Delivery Hero & https://bit.ly/3Zug150
      \\ \cmidrule{3-4}
&        & Foursquare & https://bit.ly/4io706d
      \\ \cmidrule{2-4}
& Video  &  KuaiSAR & https://bit.ly/3Vv7TzY
\\ \midrule
\multirow{3}{*}[-8mm]{Texts}    
& News         & MIND & https://bit.ly/3ZzU9Xq     \\ \cmidrule{2-4} 
& \multirow{3}{*}{Book}         & GoodReads & https://bit.ly/4f3Vjio 
      \\ \cmidrule{3-4}
&        & BookCrossing & - 
      \\ \cmidrule{3-4}
&        & WeChat-ArticleRec & -
      \\ \cmidrule{2-4} 
& Job          & Personalized & -  \\ \bottomrule
\end{tabular}
}
\label{tab:dataset}
\vspace{-2mm}
\end{table}

%% file: 7Conclusion.tex
\section{Conclusion}

Large language model-enhanced recommender systems (LLMERS) have attracted much attention due to their effectiveness and practicability. In this paper, we conclude the most recent efforts in this field. According to the component of traditional RS that LLMERS targets, we categorize the papers into three lines, \ie knowledge, interaction and model enhancement. Furthermore, to facilitate and inspire the researchers, we summarize the available resources and give out several future directions.


